\documentstyle[prl,aps,twocolumn,epsf]{revtex}
\large  

\begin{document}
\draft \twocolumn[\hsize\textwidth\columnwidth\hsize\csname
@twocolumnfalse\endcsname

\title{Diffusive persistence and the `sign-time' distribution} 
\author{T. J. Newman and Z. Toroczkai} 
\address{Department of Physics,\\ 
Virginia Polytechnic Institute and State University,\\ 
Blacksburg VA 24061,\\ USA} 
\maketitle
\begin{abstract}
We present a new method for extracting the persistence
exponent $\theta $ for the diffusion equation, based on the
distribution $P$ of `sign-times'. With the aid of a
numerically verified Ansatz for $P$ we derive an exact formula
for $\theta $ in arbitrary spatial dimension $d$. Our
results are in excellent agreement with previous numerical
studies. Furthermore, our results indicate a qualitative change
in $P$ above $d \simeq 36$, signalling the existence of a sharp
change in the ergodic properties of the diffusion field.
\end{abstract}
\vspace{5mm}  ]

\newpage
In the past few years there has been much interest
in calculating the persistence properties for
a wide range of simple model systems. Examples 
are the diffusion equation\cite{de}, the 
Ising model with Glauber dynamics\cite{ising} and
its Potts model generalization\cite{potts}, interface kinetics\cite{ik}, 
phase ordering\cite{po}, and the voter model\cite{vm}. Perhaps the
simplest and most generic system is the first -- the 
diffusion equation. Naively one might expect that everything is
known about such a classical system. However, one needs only
reflect upon its intimate relation to the rich Burgers model 
of turbulence\cite{burg}
(obtained via a simple non-linear transformation) to appreciate the potential
complexity of diffusion physics. 

This complexity was again uncovered by studies of diffusive 
persistence\cite{de}. The persistence exponent $\theta $ for this case is
defined as follows. Consider the deterministic diffusion equation
evolving a random initial condition (usually created from an uncorrelated
gaussian distribution). Then consider the probability $q(t)$ that
the diffusion field at a given site has never changed sign. One finds
numerically that this probability decays with time in a power-law fashion,
with an exponent $\theta $, whose value is {\it not} a simple rational number. 
There is no analytic prediction for $\theta $ with the exception of the 
results from the
`independent interval approximation' (IIA) which are in good agreement
with numerical work in spatial dimension $d=1$, but fare less well
in higher dimensions\cite{de}. It is easy to find applications for
diffusive persistence due to the ubiquitous presence of diffusion physics.
Examples include: survival of reactants in reaction kinetics, and 
more general survival probabilities in systems with a field slaved 
to a diffusion process.

In this Letter we shall present an exact analytic form for $\theta $.
The key to our derivation is that one may obtain $\theta $ by studying
a more general quantity; namely the distribution $P$ of `sign-times' $\tau(t)$
(to be defined below), which has also been recently introduced by
Dornic and Godr\`eche\cite{dg}. This distribution may be shown to have an exact
scaling form for all $t$: $P \ d\tau = f(\tau /t) (d\tau/t)$.
Numerically the scaling function $f$ is found to be extremely simple. Using 
this form of $f$ as an Ansatz, allows an exact determination of $\theta $. 
Our prediction deviates by $3\%$ from the numerically determined value in
$d=1$, but lies well within the error bars of the simulation results in
$d=2$ and $d=3$. We shall discuss the reasons for the 
slight $d=1$ discrepancy
toward the end of the Letter. We also find that the qualitative nature of
$P$ goes through a sharp transition at $d_{c} \simeq 36$. 

We consider the evolution of a scalar field $\phi ({\bf r},t)$ which
satisfies
\begin{equation}
\label{diff}
\partial_{t}\phi = D \nabla ^{2} \phi \ ,
\end{equation}
with initial condition $\phi ({\bf r},0) = \psi ({\bf r})$, where
the field $\psi $ is an uncorrelated random variable described by 
a gaussian distribution 
\begin{equation}
\label{distri}
R[\psi] \sim \exp \left [ -(1/2\Delta)\int d^{d}r \ \psi({\bf r})^{2} 
\right ] \ .
\end{equation}
The solution of Eq.(\ref{diff}) has the form
\begin{equation}
\label{sol}
\phi ({\bf r},t) = \int d^{d}r' \ g({\bf r}-{\bf r}',t) \ \psi({\bf r'}) \ ,
\end{equation}
where $g ({\bf r},t) = (4\pi Dt)^{-d/2} \ \exp(-r^{2}/4Dt) \ $ is the heat 
kernel.

The most obvious way to approach the persistence problem is to directly 
calculate the probability that the field at a given site (${\bf r}={\bf 0}$ 
say) has never changed sign. This amounts to the evaluation of 
\begin{equation}
\label{direct}
q(t) = \left \langle \prod \limits _{t'=0}^{t} \theta (\phi ({\bf 0},t'))
\right \rangle _{R} \ ,
\end{equation}
where $\theta(z)$ is the Heaviside step function\cite{jj}. Apart from the
IIA (which is difficult to systematically improve) 
there does not seem to be any possibility of calculating this
average. 

Let us now focus our attention on `sign-times' $\tau (t)$ defined as 
\begin{equation}
\label{st}
\tau (t) = \int \limits _{0}^{t} dt' \theta (\phi ({\bf 0},t')) \ , 
\end{equation}
so that $\tau (t)/t$ is the fraction of time $t$ in which the field at
${\bf r}={\bf 0}$ was positive. Note that this is a much easier object to
handle than $q(t)$ since we do not require the positivity of $\phi $ 
for a continuous succession of times. We simply follow the evolution
of $\phi $ and record how often it is positive. Now, the sign-time $\tau (t)$
is actually a functional of the random variable $\psi $, and as such is
described by some probability distribution $P[\tau, t]$ 
(see also Ref.\cite{dg}). We can immediately
list a few properties of $P$. First, it is defined in the domain $[0,t]$.
Second, it is symmetric about $\tau = t/2$ (since it does not matter
whether we study how often $\phi $ is positive, or how often $\phi $ is 
negative). Third, and most important, the behaviour of $P$ near $\tau = 0$ 
or $t$ directly furnishes us with the exponent $\theta $. This is clear, since
the probability for $\tau $ to be in the vicinity of either $0$ or $t$ 
is nothing more than $q(t)$ defined above. 

We refer the reader to Appendix A in which we prove an important fourth 
property of $P$. Namely, that it
assumes the scaling form $P[\tau ,t] \ d\tau = f(\tau/t) (d\tau/t)$ 
for {\it any} non-zero time $t$. [Note that $P$ does not depend on
the model parameters $D$ and $\Delta $.] 
Given the symmetry of $P$ about $\tau = t/2$, we may rewrite the 
scaling form as
\begin{equation}
\label{scaling}
P[\tau ,t] \ d\tau = g \left ( x(1-x) \right ) dx \ ,
\end{equation}
where $x=\tau/ t$ and we have assumed $f$ to be analytic around $x=1/2$. 
The third property listed above imposes that 
$g(y) \sim y^{\theta -1}$ for $y \ll 1$, where $y=x(1-x)$. 
So we have reduced the persistence problem to that of
calculating the tail of the `sign-time' distribution. This is still a
formidable task, as it involves calculating arbitrarily high moments
of the distribution (see Appendix A). 
As an alternative strategy we make an Ansatz. Namely, that the small-$y$ form
for $g$ actually holds for all $y$ in the available range $y \in [0,1/2]$.
This very simple Ansatz was both suggested and confirmed to us by the 
results of our numerical work, which we now briefly describe. (Note also
that this Ansatz appears naturally within the framework of the IIA\cite{dg}.)

Following previous work\cite{de}, we model the diffusion equation
by a discrete space-time process
\begin{equation}
\label{disc}
\phi _{i}(t+1) = \phi _{i}(t) + a\sum \limits _{j} 
[\phi _{j}(t)-\phi_{i}(t)] \ ,
\end{equation}
where the sum is over nearest neighbours of $i$ on a $d$-dimensional hypercubic
lattice. The parameter $a$ is chosen to be $1/(2d)$.
The initial value of each $\phi _{i}(0)$ is drawn from a gaussian distribution.
Simulations are performed on large lattices ($N \sim 2^{20}$ sites) for times
up to $t \sim 2^{10}$ with several independent runs. We measure two quantities
during the simulation. First, we record the number $n(t)$ of those sites 
at which the field has never changed sign. The ratio of 
$n/N \sim t^{-\theta}$ (due to the self-averaging of the system). We also
record the sign-time for each site and thus construct the histograms for
$P[\tau, t]$. Our results for the former quantity are in agreement with 
those of Ref. \cite{de}. In Figs. 1 and 2 we show the
histograms for $d=1$ and $d=2$ respectively. The results
are plotted on a log-log scale, since the data and the Ansatz are 
{\it indistinguishable} on a linear scale. One sees that the scaling function
$g$ indeed varies as a simple power law in $y$ over the range $y \in [0,1/2]$.
The only region in which there is a deviation from this power-law behaviour 
is near $y=0$. This
is purely a (time)-lattice effect, since the histogram has a finite number
of bins given by the number of time-steps. As $t$ increases, more and more
bins are supplied near $x=0$ (where the function is integrably singular)
and the expected power-law behaviour stretches closer and closer to the
origin. We stress that the power-law behaviour near $x=0$ is guaranteed by
the existence of the persistence exponent $\theta$. Our {\it a priori}
knowledge of $g$ is limited in precisely that region of $x$ 
($\in [\varepsilon ,1/2]$) where the numerical results show a clear single
power-law behaviour. We have added straight-lines on the plots, which
have slopes of $\theta -1$, with values of $\theta $ taken from 
Eq.(\ref{mu2}) below.

\begin{figure}[tbp]
\centerline{\epsfxsize=7.0cm
\epsfbox{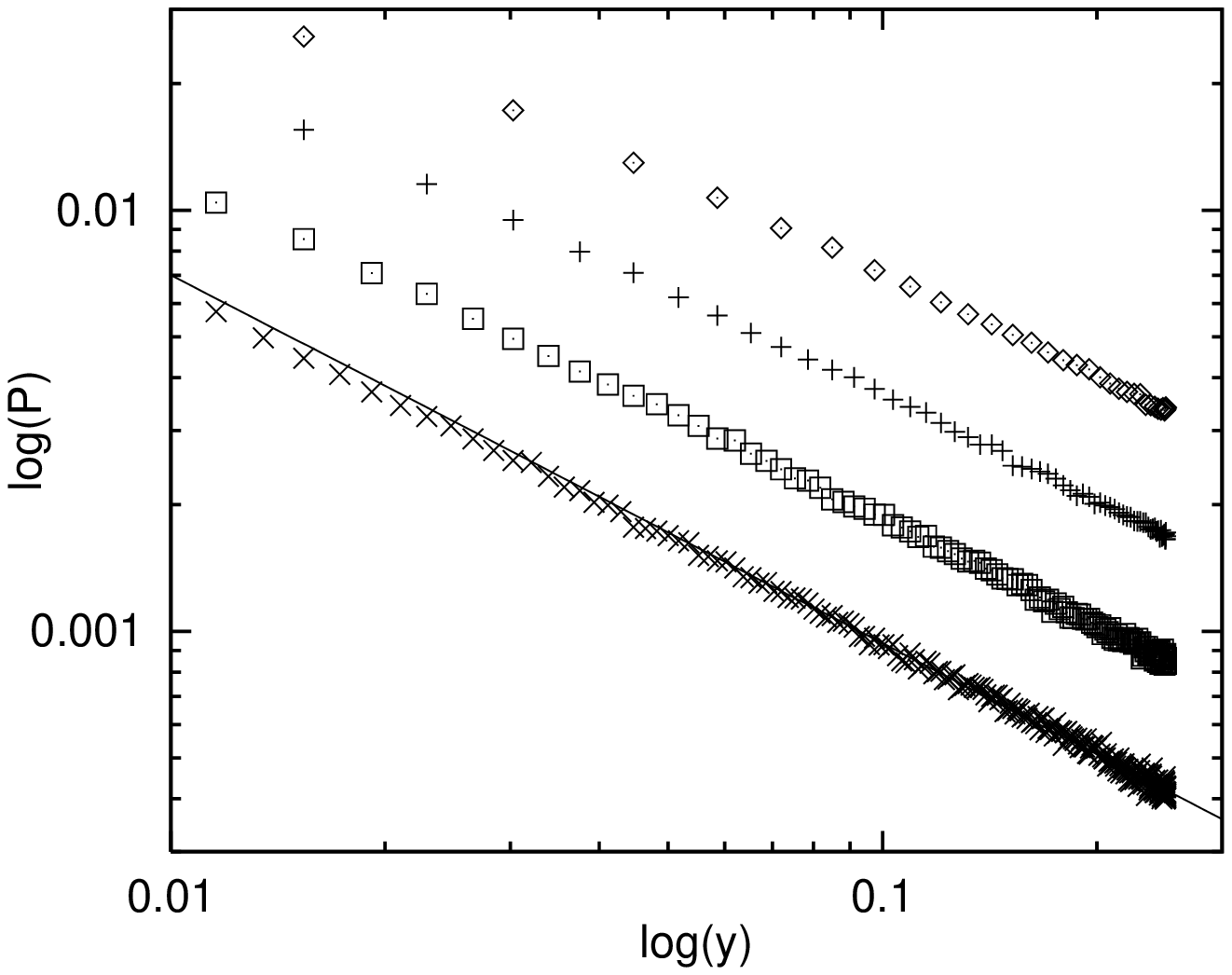}}
\vspace{0.4cm}
\caption{A log-log plot of the sign-time distribution $P$ against
$y$ for $d=1$. The data are taken at times $t=2^{n}$ with $n=6,7,8,9$
from top to bottom. The straight line is a power law with slope
$\theta -1$, $\theta = 0.1253$.}
\end{figure}

\begin{figure}[tbp]
\centerline{\epsfxsize=7.0cm
\epsfbox{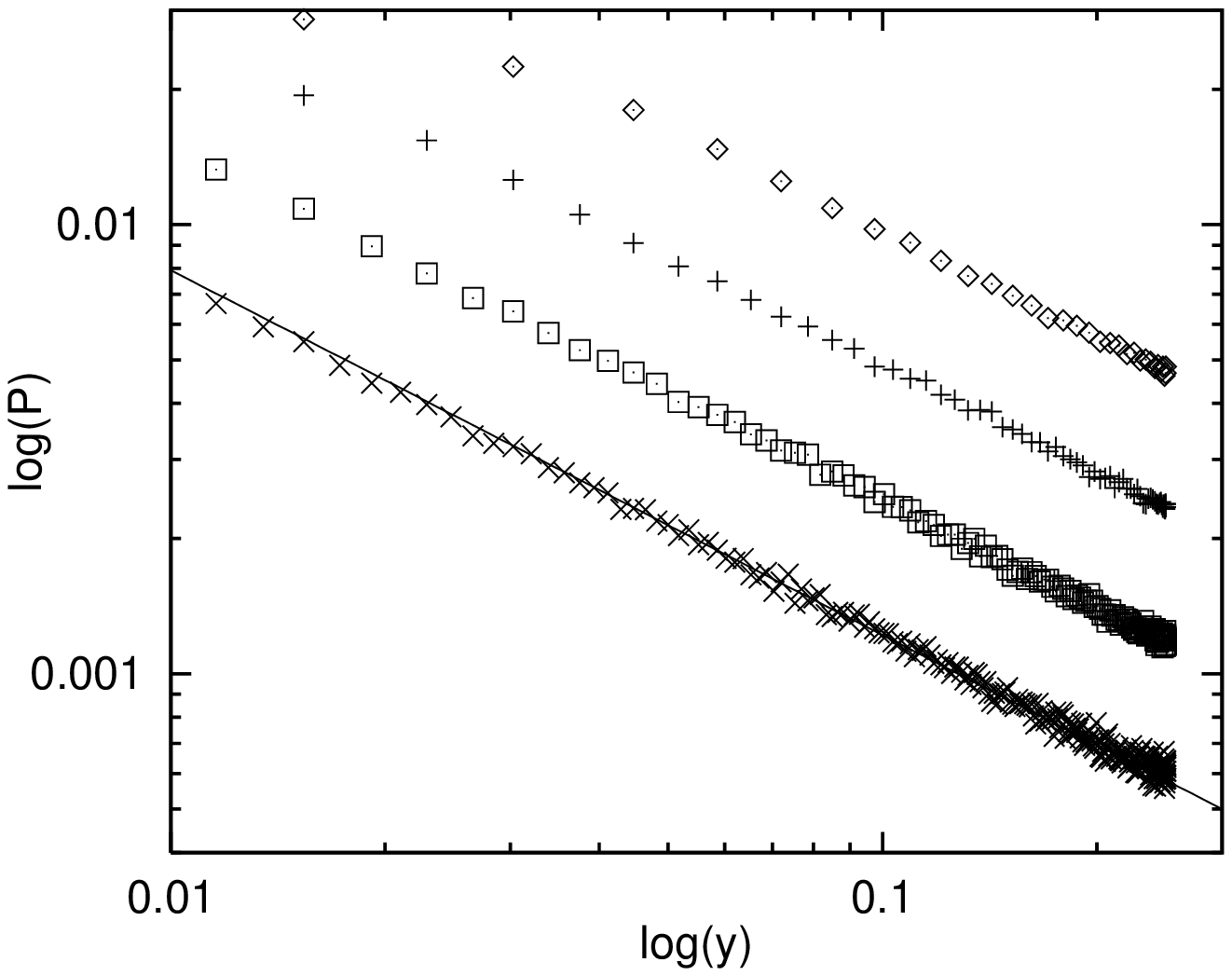}}
\vspace{0.4cm}
\caption{As above, with $d=2$ and $\theta = 0.1879$. }
\end{figure}

We are now in a position to calculate $\theta $. The numerical results
clearly support $P[\tau, t] = (c/t)[x(1-x)]^{\theta -1}$. This distribution
contains only two parameters: an amplitude $c$ (to be set by normalization),
and the exponent $\theta $. The latter can in principle be set by
the calculation of any even moment of $P$ (since the odd moments contain
no new information, as the odd cumulants are zero). The simplest to 
consider is obviously the second moment $\mu _{2}$. (Unfortunately
it is extremely difficult to calculate any even moments above the second.) 
The integrals over $P$ are
simply given in terms of the Beta function\cite{as}. From the normalization
one fixes $1/c = B(\theta, \theta)$. The calculation of the second
moment yields 
\begin{equation}
\label{smom}
\mu _{2} = {B(\theta+2,\theta) \over B(\theta,\theta)} = 
{(1+\theta)\over 2(1+2\theta)} \ .
\end{equation}

We may now independently calculate $\mu _{2}$ from the original definition
of the sign-times. Explicitly we have:
\begin{equation}
\label{mom2}
\mu_{2} = \langle (\tau (t)/t)^{2} \rangle _{R} = \int \limits _{0}^{1} da_{1}
\int \limits _{0}^{1} da_{2} \ C_{2}(a_{1}t,a_{2}t) \ ,
\end{equation}
where
$C_{2} = \langle \theta (\phi({\bf 0},a_{1}t)) \theta (\phi({\bf 0},a_{2}t))
\rangle _{R} \ .$
This latter quantity may be calculated exactly to give 
(see Appendix for a hint)
\begin{equation}
\label{cor2ex}
C_{2} = {1\over 4} \ + \ {1\over 2\pi} {\rm sin}^{-1} 
\left [ \left ( {2a_{1}^{1/2}a_{2}^{1/2}\over (a_{1}+a_{2})} \right ) 
^{d/2} \right ] \ .
\end{equation}
The integrals in (\ref{mom2}) are easily performed, leaving one with 
the expression
\begin{equation}
\label{mu2}
\mu_{2}={1\over 4}(2-\beta) \ \ \ \Rightarrow \ \ \ \theta = {\beta \over 
2(1-\beta)} \ ,
\end{equation}
where 
\begin{equation}
\label{beta}
\beta (d) = {d \over 2\pi} \int \limits _{0}^{1} da \
{(1-a) \over (1+a)} \left [ \left ( {1+a\over 2a^{1/2}} \right ) ^{d} -1
\right ] ^{-1/2} \ .
\end{equation}
This integral may be performed explicitly in one and two dimensions,
with the results:
\begin{eqnarray}
\label{bet1}
\nonumber
\beta (1) & = & {1\over 2^{3/2} \pi} \bigl  [ \psi (11/8) + \psi (9/8) 
- \psi (7/8) - \psi (5/8) \bigr ] \\
& & \ \ \ \ = 8\sqrt{2}/3\pi \ - \ 1 = 
0.20042\cdots \ ,
\end{eqnarray}
and 
\begin{eqnarray}
\label{bet2}
\nonumber
\beta (2) = {1\over \pi} \bigl [ \psi (5/4) & - & \psi (3/4) \bigr ] \\ 
& = & 4/\pi \ - \ 1 = 0.27323\cdots \ ,
\end{eqnarray}
where $\psi (z)$ is the digamma function\cite{as}, showing that the persistence
exponent is {\it not} a simple rational number, but is transcendental.
We refer the reader to Table 1 where values of $\theta (d)$
are listed, along with the numerical and IIA estimates of Ref.\cite{de}. 

It is of interest to calculate the large-$d$ form for $\theta $.
Then one finds:
\begin{equation}
\label{larged}
\theta (d) = {\pi \over 4I} \left ( {d \over 2} \right ) ^{1/2} + O(1) \ ,
\end{equation}
where 
\begin{equation}
\label{largedex}
I = \int \limits _{0}^{\infty} dw \ {[\log (1+w)]^{1/2} \over w^{1/2}(1+w)}
\ = 3.0005 \cdots \ ,
\end{equation}
giving a value of $\theta (d) \sim (0.1850\cdots )\sqrt {d}$.
This is to be compared to the result from the IIA which gives 
$\theta _{\rm IIA} (d) \sim (0.1454\cdots )\sqrt {d}$. As can be seen from 
this large-$d$ limit, and also from Table I, the IIA consistently
underestimates the value of $\theta $.

The fact that $\theta $ passes through unity is very interesting as
it has a direct consequence for the sign-time distribution $P$. For
$\theta < 1$, $P$ has integrably divergent tails at $\tau (t) \rightarrow
0$ and $\tau (t) \rightarrow t$. Also, the mean of the $P$ (which is
at $\tau =t/2$) is the {\it least} likely value of $\tau $. One can understand
this by considering a given point being located in the centre of
a very large positive domain. A long time must pass before a negative
domain sweeps through, thus halting the sign-time clock. However, if
$\theta >1$, the distribution becomes convex, and the tails go to
zero at the end-points of the range $[0,t]$. Thus the mean value
$\tau = t/2$ is in this case the {\it most} likely value of $\tau $. 
The dynamic mixing of positive and negative domains is in this case
very efficient. (We shall return to this point in our conclusions.)
It is therefore important to know at what dimension $\theta $ passes
through unity. This is equivalent to insisting that $\beta (d_{c}) = 2/3$.
Numerical evaluation of the integral in Eq.(\ref{beta}) using Mathematica,
yields the value $d_{c} = 35.967 \cdots $. The fact that this enormous
dimension plays a physical role in diffusion physics is extraordinary at
first sight: such is the complexity of diffusive persistence.

\begin{center}
\begin{tabular}{|c||c|c|c|}\hline
{$d$} & {$\theta$} & {$\theta _{\rm IIA}$} & {$\theta _{\rm sim}$} \\ \hline 
  1 & $0.1253\cdots$ & $0.1203\cdots$ & 
\ \ $0.1207(5) $ \ \ \ \\ \hline
  2 & $0.1879\cdots$ & $0.1862\cdots$ & 
\ \ \ $0.1875(10) $ \ \ \ \\ \hline
  3 & $0.2390\cdots$ & $0.2358\cdots$ & 
\ \ \ $0.2380(15) $ \ \ \ \\ \hline
 \ \ $\gg 1$ \ \ & \ \ $0.1850\cdots \sqrt {d}$ \ \ & 
\ \ $0.1454\cdots \sqrt {d}$ \ \ & -- \\ \hline 
\end{tabular}
\end{center}

\vspace{0.2cm}

\noindent
Table 1: The calculated value of $\theta $ from Eqs.(\ref{mu2}) and 
(\ref{beta}), along with the IIA and simulation estimates from Ref.\cite{de}.

\vspace{0.2cm}

Before concluding, we wish to make some comments regarding the
role of the space-time lattice. It is quite possible to set
up a calculation of the persistence exponent, or the sign-time
distribution, with the discrete algorithm (\ref{disc}) as a starting
point. This formulation has the advantage that the direct calculation
of $q(t)$, as defined by Eq.(\ref{direct}), is well-defined. [In the
continuum a microscopic time cut-off must be introduced to make sense
of the time slices.] We have also pursued the discrete formulation
in an attempt to calculate $P[\tau, t]$. As with the continuum case, it is 
extremely difficult to calculate any even moment above the second. However, 
we find the interesting result that the second moment contains strong
corrections to scaling; i.e. $\langle \tau (t)^{2} \rangle \sim t^{2}
+O(t)$. In the continuum, each moment has a clean scaling  
$\langle \tau (t)^{n} \rangle \sim t^{n}$ for any non-zero time. The
implication of this result, is that one should not expect true 
scaling from numerical work (based on (\ref{disc})) until very late times.
This is especially true in low dimensions where the lattice Laplacian
is much weaker than its continuum counterpart. This effect is already
apparent in the histograms shown in Figs. 1 and 2. One sees that the
tail (for small $y$) bends away from its power-law form due to the discrete
sampling on a finite time grid. As more time steps are used, this deviation
from scaling is pushed to smaller values of $y$. It should be stressed that
the direct measurement of persistence from the ratio $n(t)/N$ is less
reliable since the measurement is equivalent to sampling $P$ at $y=0$ --
precisely in the region most affected by finite time-step effects. It
is for these reasons that we believe the $d=1$ measurement of $\theta $,
as given in Ref.\cite{de}, to be the least solid. In principle one can
avoid these finite time-grid effects by sampling data from the exact
solution of the diffusion equation, as given in (\ref{sol}), where $t$
is a real, continuous quantity. Such a numerical study would require
orders of magnitude more computer time than the previous studies,
but could potentially give a definitive answer for the case $d=1$.

In conclusion we have provided an exact form for the diffusive
persistence exponent $\theta $, with the aid of an Ansatz
based on numerical observations of the sign-time distribution $P$.
Naturally, this result is in no way rigorous. We consider a proof
of our Ansatz to be a reachable goal, although such a proof
involves very technical manipulations in $n$-dimensional geometry\cite{ls,tn}.
We have also demonstrated the wider significance of the persistence exponent
in parameterizing the sign-time distribution. We believe that this
distribution will become an important new tool in other persistence-type
problems. The tails of the distribution contain standard 
persistence information,
whilst the body of $P$ gives important information regarding the
mixing efficiency of the problem at hand. If $P$ is concave (convex), than the
mean value is the least (most) likely. Such information is important
when ergodic properties of a system are under investigation. The fact that
the curvature of
$P$ for the simple diffusion equation `flips' at $d=d_{c} \simeq 36$
indicates that the dynamics in the phase space of few-body systems
may be extremely sensitive to exactly how many degrees of freedom are
considered. 

The authors are grateful to A. Bray, E. Ben-Naim, C. Godr\`eche and 
R. Zia for interesting discussions and also
thank C. Godr\`eche for bringing Ref.\cite{dg} to their
attention prior to publication.
T.J.N. and Z.T. acknowledge financial support from the Materials
Research Division of the National Science Foundation.
Z.T. also acknowledges support from the Hungarian Science Foundation,
T17493 and T19483.

\appendix

\section{}

\vspace{-0.3cm}

In this appendix we sketch a brief proof of the assertion that
$P$ satisfies the exact scaling form $P[\tau, t]d\tau = f(\tau /t)(d\tau /t)$
for all $t$.
First, we note that the sign-time distribution can be written as
$P[\tau, t] = \langle
\delta (\tau - \tau _{\psi}) \rangle _{R}$, where $\tau _{\psi}$ is the
implicit function of $\psi $ given in (\ref{st}). Thus,
\begin{equation}
\label{app1}
P[\tau, t] = \int \limits _{-\infty }^{\infty } {d\omega \over 2\pi} \ 
e^{i\omega \tau } \left \langle \exp (-i\omega \tau _{\psi}) 
\right \rangle _{R} \ .
\end{equation}
The average on the rhs of the above expression may be re-expressed as a power
series in terms of $\omega ^{n} \langle \tau _{\psi}^{n}\rangle _{R}$.  
The $n^{\rm th}$ moment is an $n$-fold integral over the average of $n$
step-functions. Each step-function may be represented by an integral,
yielding (with an implicit limit of $\epsilon _{k} \rightarrow 0$)
$$ \langle \tau _{\psi}^{n}\rangle = {1\over (2\pi )^{n}}
\prod \limits _{j=1}^{n} \int \limits_{0}^{t}dt_{k}
 \int \limits _{-\infty}^{\infty}
{d\sigma _{k}\over (\epsilon_{k}+i\sigma _{k})}
\left \langle e^{i \sum_{l} 
\sigma _{l}\phi ({\bf 0},t_{l})}\right \rangle _{R}\ .$$
The average is easily performed over the gaussian distribution $R[\psi ]$.
On rescaling the integration variables we find
$$ \langle \tau _{\psi}^{n}\rangle = \left ({t\over 2\pi}\right )^{n}
\prod \limits _{k=1}^{n} \int \limits_{0}^{1}da_{k}
 \int \limits _{-\infty}^{\infty} {d\sigma_{k}'\over 
(\epsilon_{k}'+i\sigma_{k}')}
\left [e^{-\sum _{l,m}\sigma_{l}'M_{l,m}\sigma_{m}'} \right ]$$
where $M_{l,m}=[2a_{l}^{1/2}a_{m}^{1/2}/(a_{l}+a_{m})]^{d/2}$.
We see that the $n^{\rm th}$ moment scales exactly as $t^{n}$, which enables
us to combine $\omega$ and $t$ as a simple product.
Returning to (\ref{app1}), we may scale $\omega $ by $t$ to obtain
the desired result.

\end{document}